\UseRawInputEncoding
\PassOptionsToPackage{breaklinks=true}{hyperref}
\documentclass[pra,floatfix,twocolumn,superscriptaddress,showpacs,preprintnumbers,nofootinbib,longbibliography
]{revtex4-2}
\usepackage{dcolumn}    
\usepackage{bm} 
\usepackage{graphicx}
\usepackage{lipsum}
\usepackage{amsmath}  
\usepackage{amsthm}
\usepackage{amssymb}
\usepackage{appendix}
\usepackage{enumitem}
\usepackage{latexsym}
\usepackage{amsfonts}  
\usepackage{mathtools} 
\usepackage{xcolor}
\usepackage{array}      
\usepackage{epsfig}
\usepackage{braket} 
\usepackage{bbold}
\usepackage{float}
\usepackage{color}
\usepackage{soul}
\usepackage[normalem]{ulem}
\usepackage{xurl}
\usepackage[colorlinks=true,linkcolor=blue,urlcolor=blue,citecolor=blue,pdfusetitle]{hyperref}
\UseRawInputEncoding

\begin{document}

%\title{Benchmarking dynamics against resource-theoretical predictions}
%\title{Near-Optimal Thermodynamic Performance from Realistic Open-System Environments}
\title{Approaching Resource-Theoretic Optimal Performance with Structured Environments}
%\title{Resource-theoretic Optimal Performance Can Be Achieved with Realistic Environments}
\author{\ Lea Lautenbacher}
\email[]{lea.lautenbacher@uni-ulm.de}
\affiliation{Institut f\"ur Theoretische Physik, Albert-Einstein-Allee 11, Universit\"at Ulm, D-89069 Ulm, Germany}
\author{\ Giovanni Spaventa}
\email[]{giovanni.spaventa@uni-ulm.de}
\affiliation{Institut f\"ur Theoretische Physik, Albert-Einstein-Allee 11, Universit\"at Ulm, D-89069 Ulm, Germany}
\author{\ Susana F. Huelga}
\email[]{susana.huelga@uni-ulm.de}
\affiliation{Institut f\"ur Theoretische Physik, Albert-Einstein-Allee 11, Universit\"at Ulm, D-89069 Ulm, Germany}
\affiliation{Center for Integrated Quantum Science and Technology (IQST), 89081 Ulm, Germany}
\author{\ Martin B. Plenio}
\email[]{martin.plenio@uni-ulm.de}
\affiliation{Institut f\"ur Theoretische Physik, Albert-Einstein-Allee 11, Universit\"at Ulm, D-89069 Ulm, Germany}
\affiliation{Center for Integrated Quantum Science and Technology (IQST), 89081 Ulm, Germany}

\begin{abstract}

Resource-theoretic approaches to thermodynamics provide powerful, model-independent bounds on the efficiency of physical processes, because they do not rely on microscopic details of the environment. Whether such bounds can be approached by realistic dynamics generated by explicit system-environment interactions remains an open question. Photoisomerization, a fundamental molecular photoreaction, offers a concrete setting to examine this issue. We introduce a tunable microscopic model of a molecular photoswitch coupled to a structured vibrational environment, which interpolates continuously between Markovian and non-Markovian regimes. Resource-theoretic analysis predicts in particular that Markovian Thermal Operations achieve strictly lower yields than general Thermal Operations. We show that environmental memory lifts dynamical restrictions associated with Markovian thermal evolutions, thereby enlarging the set of transformations accessible to the microscopic dynamics. Approaching the thermal operation bound, however, depends on the microscopic coupling structure that generates this memory and directs the resulting dynamics towards the target transformation.

\end{abstract}

\maketitle

\section{Introduction}
\label{Intro}

Assessing the efficiency of physical processes is a central problem in physics. In the quantum regime, nonclassical resources such as entanglement and coherence offer pathways to surpass classical performance limits \cite{PlenioV07, Martin17}. Quantum resource theories (QRTs), developed within quantum information science, provide a rigorous framework for quantifying and comparing such resources without specifying the microscopic details of the underlying dynamics \cite{Coecke, Chitambar, GourBook}.

Resource theories of thermodynamics have emerged as a powerful framework for characterizing the constraints and capabilities of physical processes that are consistent with the second law of thermodynamics \cite{Ruch76, Horodecki13, Goold16, Nelly, Brandao, LostaglioReview} while extending to situations far from equilibrium and at the nanoscale for which traditional thermodynamics does not apply. 
In this framework, efficiencies are assessed under thermal operations (TOs), which require no external work input and rely solely on interaction with a thermal bath. Crucially, the resulting bounds on maximal achievable efficiency are fully independent of the microscopic structure of the environment \cite{Giovanni, Nicole, tiwary2024quantum, siciliano2025correlations, burkhard2024boosting}.

In parallel, extensive work has sought to elucidate how environmental interactions shape transport and energy-conversion in open quantum systems \cite{Deph_assisted,mohseni2008environment,caruso2009highly,Somoza}. These efforts span systems ranging from engineered solid-state devices to biological complexes \cite{Biobook}, where the microscopic structure of the environment is often only partially known yet exerts a decisive influence on the dynamics \cite{HuelgaP2013}. Photoisomerization exemplifies this situation: it is a fundamental biological process whose efficiency is central to vision and light sensing, yet its microscopic foundations remain incompletely understood. Although sophisticated nonadiabatic and ab initio studies of retinal and related systems exist, no single, fully characterized microscopic Hamiltonian captures the full range of biologically relevant environmental interactions \cite{Orlandi_07, Domcke94}. As a result, theoretical modeling necessarily relies on simplifying assumptions and focuses on selected dynamical aspects.

In this paper, we investigate whether environmental memory can relax the dynamical restrictions associated with Markovian thermal evolutions, thereby allowing a microscopic model to access transformations that would otherwise remain dynamically inaccessible. Earlier work \cite{Giovanni, Nicole, tiwary2024quantum, siciliano2025correlations, burkhard2024boosting} has shown that QRTs can yield upper bounds on the achievable photoisomerization yield, revealing both the ultimate thermodynamic limit set by general thermal operations and the tighter constraints imposed by Markovian thermal operations (MTO) \cite{Giovanni}, a strict subset of TO. Despite their rigor, resource-theoretic approaches are often regarded as detached from concrete physical dynamics, especially in complex molecular or biological contexts where a complete microscopic Hamiltonian is unknown. Our aim is therefore to assess whether these resource-theoretic model-independent bounds remain informative for realistic dynamical models, and to identify the microscopic mechanisms that enable, or preclude their saturation.

We address this question within a tunable microscopic model of a molecular photoswitch coupled to a structured vibrational environment described by the pseudomode formalism \cite{Garraway, Imamoglu, Lemmer, somoza2019dissipation, Nicola24}. This construction enables a smooth interpolation between Markovian and non-Markovian regimes while introducing neither external work nor coherence, ensuring compatibility with the thermodynamic assumptions underlying TO and MTO. Comparing the resulting dynamics with the corresponding resource-theoretic bounds reveals the conditions under which non-Markovianity enables efficiencies that exceed those achievable by Markovian thermal operations, as well as structural features of the dynamics that limit this enhancement. More broadly, this framework offers a resource-theoretic perspective on earlier observations that coherent coupling to selected vibrational modes can enhance the performance of microscopic energy-transfer models~\cite{killoran2015}, placing such effects in the context of dynamical accessibility under thermodynamic constraints.

The paper is organized as follows. In Sec. II we briefly review the relevant aspects of thermodynamic resource theories and the bounds they impose on photoisomerization  yields. In Sec. III we introduce our microscopic model of a molecular photoswitch coupled to a structured vibrational environment. Sections IV and V present the results of our dynamical simulations. In Sec. VI we analyze the set of states accessible to the dynamics and compare it to the cones defined by resource-theoretic operations. We summarize our conclusions and discuss the broader implications in Sec. VII.

\section{Thermodynamic resource theories}

The resource-theoretic framework specifies a set of allowed “free” states and operations, subject 
to given constraints, and characterizes the possible state transformations enabled by these free 
operations \cite{Coecke, Chitambar,GourBook}. These theories emphasize what transformations are 
achievable rather than describing the explicit dynamics generating them. This leads to an input-output 
perspective on state conversion and clarifies the operational consequences associated with different
resources.

In thermodynamic resource theories, the free state is the Gibbs state, and the largest class of free operations consists of all channels that preserve it, known as Gibbs-preserving (GP) operations. In this work, we focus on a subclass of these operations: the thermal processes (TPs), which are free operations additionally constrained by thermomajorization \cite{LostaglioReview}. A TP is described by a quantum channel 
$\mathcal{E}$ that satisfies two main properties:

(P1) \textit{Stationary thermal state}. The Gibbs state 
\begin{eqnarray}
    \tau &=& \frac{e^{-\beta H_S}}{\text{Tr}(e^{-\beta H_S})}, 
\end{eqnarray}
where $\beta = 1/k_{B}T$ is the inverse temperature of the bath, and $H_S$ is the system Hamiltonian,
is a fixed point of the operation, 
\begin{eqnarray}
    \mathcal{E}(\tau) &=& \tau.
    \label{eq:gibbs}
\end{eqnarray}

(P2) \textit{Phase covariance/Time translation symmetry}. The channel $\mathcal{E}$ commutes with the unitary time evolution $\mathcal{U}$ 
\begin{equation}
\label{eq:PC}
\mathcal{E}(\mathcal{U}(\rho)) = \mathcal{U}(\mathcal{E}(\rho)),
\end{equation}
or equivalently $\mathcal{E}\,\circ\, \mathcal{U} =\mathcal{U}\, \circ\, \mathcal{E}$, where $\mathcal{U}(.) = e^{-i H_S t} (.) e^{i H_S t}$. Such a map $\mathcal{E}$ is known as a \textit{phase covariant} map. 

In the open system scenario, phase covariance can be established by imposing the rotating wave approximation (RWA). The presence of counter-rotating terms in the interaction Hamiltonian does not conserve energy with respect to $H_S+H_E$. Under RWA these rapidly oscillating terms are neglected. As a consequence, imposing RWA not only simplifies the mathematical structure of the system-environment interaction but also enforces a dynamical constraint that yields phase covariance.

Another class of free operations is the class of TOs, such that $\text{TO}\subseteq \text{TP} \subseteq \text{GP}$. A thermal operation $\mathcal{T}$ is induced by unitary interactions between the system $S$ and an environment, a thermal bath $B$, corresponding to a thermal state at an inverse temperature $\beta$ defined as 
\begin{equation}
    \label{eq:Tmap}
    \mathcal{T}(\rho) = \text{Tr}_B\left[ U (\rho_S \otimes \tau_B)U^\dagger\right], 
\end{equation}
where the unitary $U$ satisfies energy conservation encoded via $[U, H_S + H_E] = 0$, with $H_S$ and $H_E$ the Hamiltonians of system and bath, respectively. The condition of strict energy preservation ensures that the principles of thermodynamics hold even at the nanoscale, where quantum effects and boundary interactions play a significant role. This condition captures all processes that can be realized without an external source of work, and consequently populations and coherences evolve independently. There are no further constraints on $U$, hence even strong correlations between system and bath may build up, and one can expect the presence of non-Markovian effects. Under the additional constraint of Markovianity we arrive at the class of Markovian thermal operations, a subset of TOs generated by Gibbs-preserving, time-inhomogeneous Lindblad dynamics. In the same way that state conversion under TOs is captured by thermomajorization \cite{Horodecki13}, state conversion under MTO is captured by the stricter, continuous, notion of Markovian thermomajorization which encapsulates all constraints that memoryless thermal processes impose on population dynamics \cite{LostaglioMark,Giovanni}. 

\section{The model}

\subsection{Molecular switches}

Resource theories have proved to be powerful tools to compute general bounds on the efficiency of processes without the need of a proper microscopic description of the process.
A molecular switch, and in particular a photoisomer, is a molecule that, upon absorbing light in a process known as photoabsorption, undergoes a structural change. This transformation typically involves the physical rotation of certain chemical groups within the molecule relative to others, rather than a rotation of the entire molecule as a rigid body. This intramolecular reconfiguration, often around a specific bond, leads to the \textit{cis-trans} isomerization that underlies the switching behavior. Note that no chemical bond is broken in the process. In this work, we describe the molecule by a 4-level system representing the electronic eigenstates,  $|e_0\rangle, |e_1\rangle, |e_2\rangle$ and $|e_3\rangle$, with energies $E_0, E_1$, $E_2$ and $E_3$, respectively. The free Hamiltonian of the system is given by 
\begin{equation}
\label{eq:HS}
H_S = \sum_{i=0}^3 E_i|e_i\rangle \langle e_i| \, .
\end{equation} 
Considering the state of the system initially incoherent and making use of the fact that thermal operations do not mix populations and coherences, it can be equivalently described by its population vector of $\rho_S (0)$
\begin{equation}
\label{eq:rhoSini}
\mathbf{p}_0 = (1-q,q,0,0), 
\end{equation}
where $q \in [0,1]$ denotes the photoexcitation parameter. Aiming to analyze how population is transferred, efficiency can be defined as the amount of population in the desired electronic state after the process, given by the yield $\gamma$ as  
\begin{equation}
    \label{eq:yield}
    \gamma = \langle e_2|\rho_S (t)|e_2\rangle \, . 
\end{equation}
In the following dynamical analysis, $t$ denotes the chosen output time of the process. The optimal yield $\gamma_{\rm TO}$, obtained after optimizing the equation above over all possible thermal operations is given by \cite{Giovanni}
\begin{equation}
    \label{eq:opt_yield}
    \gamma_{\rm TO} = 
    \begin{cases}
        q +(1-q)(e^{-\beta E_2} - e^{- \beta E_1}),& q\geq \Tilde{q}\\[0.1cm]
        (1-q)e^{- \beta E_2},            & q<\Tilde{q},
    \end{cases}
\end{equation}
where $\Tilde{q} = 1/(1+e^{\beta E_1})$. The same optimization for Markovian thermal operations yields
\begin{equation}
    \label{eq:mark_yield}
    \gamma_{\rm MTO} = 
    \begin{cases}
        [q +(1-q)\frac{e^{- \beta E_2}}{1+e^{-\beta E_2}}]\frac{e^{-\beta E_2}}{e^{-\beta E_2} + e^{- \beta E_1}},& q\geq \Tilde{q}\\[0.2cm]
        [1-q \frac{e^{-\beta E_1}}{e^{-\beta E_2} + e^{-\beta E_1}}]\frac{e^{-\beta E_2}}{1+e^{-\beta E_2}},            & q<\Tilde{q}.
    \end{cases}
\end{equation}
As previously studied and discussed \cite{Giovanni}, the optimal yield Eq.~\eqref{eq:opt_yield} cannot be achieved under the additional restriction of Markovianity. We emphasize that both optimal $\gamma_{\rm TO}$ and Markovian $\gamma_{\rm MTO}$ bounds were obtained under an optimization over all possible TOs and MTOs respectively, i.e. over all possible, and potentially highly complex, environments. This raises an issue regarding the achievability of these bounds for specific, physically realistic environments, typical of biomolecular complexes.

\begin{figure}[t]
    \centering
    \includegraphics[width=\linewidth]{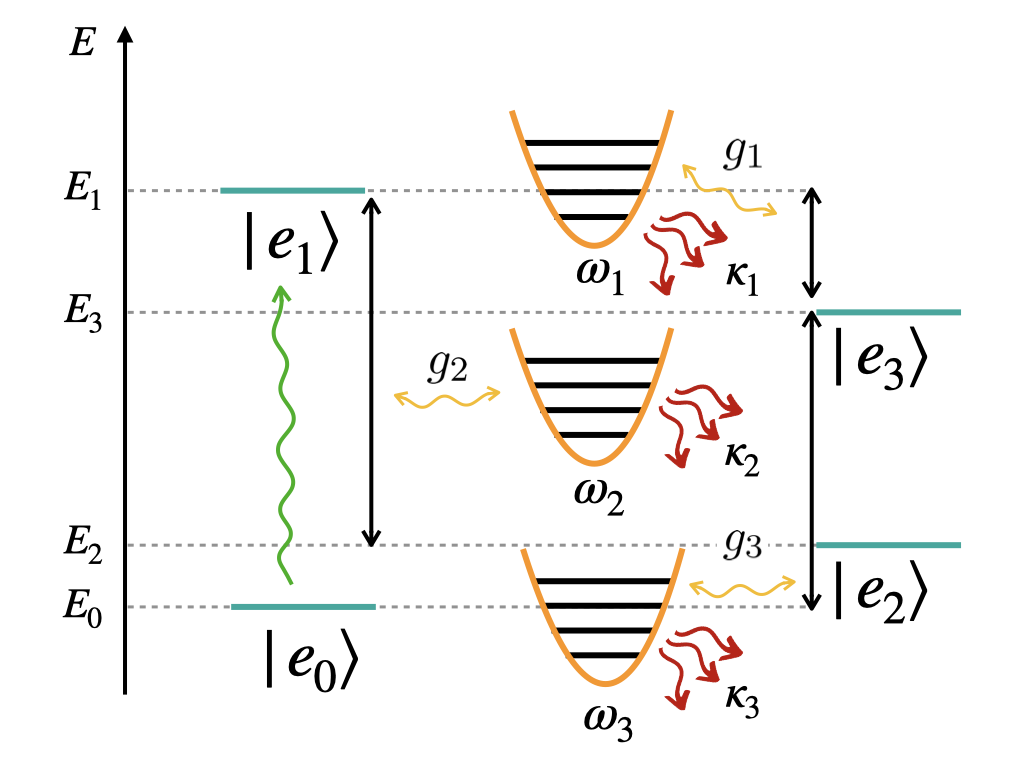}
    \caption{Minimal model of a photoisomer coupled to a structured vibrational environment. A four-level electronic system interacts with three pseudomodes with coupling strengths $g_i$. Each pseudomode is damped into an independent Markovian reservoir at rate $\kappa_i$. The parameters $g_i$ and $\kappa_i$ control the transition between Markovian and non-Markovian regimes.}
    \label{fig:model}
\end{figure}

\subsection{Microscopic modeling}
\label{sec:model}

To investigate the achievability of these bounds for a system in contact with an explicit environment, we propose a microscopic quantum model to understand the mechanism of energy and electron transfer processes at the molecular level. Our aim here is to construct a minimal but tunable model that captures essential features of vibrationally driven photoisomerization  while allowing a systematic comparison with the bounds set by resource theories. The pseudomode formalism provides a controlled way to introduce structured vibrational environments \cite{Garraway, Imamoglu, Lemmer, somoza2019dissipation, Nicola24} and to interpolate between the Markovian and the non-Markovian regime \cite{NM_assisted}. We consider a structured reservoir, modeled by three pseudomodes, that couples to the electronic transitions and ultimately drives the system's configuration cis-trans transition. A schematic representation of the model is provided in Fig.~\ref{fig:model}. 

The photoisomer is modeled by the same electronic system described by $H_S$ of Eq.~\eqref{eq:HS}. The system is in contact with a vibrational environment, described here as a collection of damped bosonic modes, the pseudomodes, with free Hamiltonian 
\begin{equation}
    \label{eq:HM}
    H_{E} = \sum_{i=1}^{3}\omega_{i}a^{\dagger}_{i}a_{ i} \, ,
\end{equation}
where $\omega_{i}$ is the mode frequency, $a^{\dagger}_{i}$ the creation and $a_{i}$ the annihilation operators of the modes. The electronic system interacts with the vibrational environment via the interaction Hamiltonian under rotating wave approximation 
% \begin{equation}
% \label{eq:H_int}
% H_I = \sum_{i=0}^3 g_i \left( |e_{\alpha_i}\rangle \langle e_{\beta_i}|\, a_i + \mathrm{h.c.} \right),
% \end{equation}
% where the $g_{i}$ are the coupling constant of each electronic transition and the frequency of each corresponding mode matches that of the electronic transition $\omega_i = E_{\alpha_i} - E_{\beta_i}$ with 
% \begin{equation}
% \label{eq:transitions}
% (\alpha_i, \beta_i) = 
% \begin{cases}
% (1,3), & i=1 \\
% (1,2), & i=2 \\
% (3,0), & i=3\,\, .
% \end{cases}
% \end{equation}
\begin{equation}
\label{eq:H_int}
H_I = \sum_{i=1}^{3} g_i \left( A_i\, a_i^\dagger + A_i^\dagger a_i \right),
\end{equation}
where the operators $A_i$ describe the electronic transitions
\begin{equation}
\label{eq: A_ops}
A_1 = |e_3\rangle\langle e_1|, \quad 
A_2 = |e_2\rangle\langle e_1|, \quad 
A_3 = |e_0\rangle\langle e_3|.
\end{equation}
The frequency of each bosonic mode matches the corresponding electronic transition to enforce energy conservation of the interaction 
\begin{equation}
\omega_1 = E_1 - E_3, \qquad
\omega_2 = E_1 - E_2, \qquad
\omega_3 = E_3 - E_0 .
\end{equation}
Note that the form of the interaction in Eq.~\eqref{eq:H_int} is chosen such that, under the resonance condition, it conserves the total bare energy, (see Appendix~\ref{ap:modes_thermal}). Together with the thermal character of the environment, this places the corresponding input-output transformations within the thermodynamic setting underlying thermal operations. The model can therefore be viewed as a restricted microscopic realization that provides a natural baseline for comparison the model-independent TO performance bound. The total Hamiltonian is 
\begin{equation}
\label{eq:Htot}
    H = H_S + H_E + H_{I} \, . 
\end{equation}
The initial state of the system $\rho_S(0)$ is parametrized as in Eq.~\eqref{eq:rhoSini} and the modes are in the thermal equilibrium state $\tau_i$, i.e. the Gibbs state at inverse temperature $\beta$. The composite initial state of the system and vibrational environment is
\begin{equation}
\rho(0) = \rho_S(0)\otimes \bigotimes_{i=1}^{3}\, \tau_{i} \, . 
\end{equation}
The global evolution of electronic system and vibrational modes, is described by a Lindblad-type master equation of the form
\begin{equation}
\label{eq:lind}
\dot{\rho} = -i[H,\rho(t)] + \mathcal{D}(\rho(t))
\end{equation}
where each mode dissipates to its own independent Markovian 
environment as described by the dissipator 
\begin{equation}
    \label{eq:diss}
    \mathcal{D}(\rho) = \sum_{i=1}^3 \kappa_i(1+\bar{n}_i)\,D[a_i](\rho) +\kappa_i\bar{n}_i\,D[a_{i}^{\dagger}](\rho)\, , 
\end{equation}
where 
\begin{equation}
\label{eq:diss2}
D[L](\rho)=L\rho L^\dagger -\frac{1}{2}\{L^\dagger L,\rho\}\,,
\end{equation}
for a jump operator $L$. Here $\kappa_i$ is the damping rate of the $i-$th mode, and $\bar{n}_i$ is the mean number of excitations given by the Bose-Einstein distribution
\begin{equation}
    \label{eq:nbar}
    \bar{n}_i = \frac{1}{e^{\beta \omega_i } - 1} \, . 
\end{equation}
We define Markovianity in terms of the divisibility properties of the dynamical map \cite{Rivas10}. Specifically, let
$\{\mathcal{E}_{t, t_0}\}_{t\geq t_0}$ be a family of CPTP maps describing the reduced evolution of the electronic system. The evolution is said to be \textit{CP-divisible} if, for every $t\geq s\geq t_0$, there exists a completely positive and trace-preserving intermediate map $\mathcal{E}(t,s)$ such that
\begin{equation}
\mathcal{E}(t,t_0)=\mathcal{E}(t,s)\,\mathcal{E}(s,t_0).
\end{equation}
Throughout this work, we adopt CP-divisibility as our criterion of quantum Markovianity: CP-divisible evolutions are referred to as \textit{Markovian}, whereas a failure of CP-divisibility signals non-Markovian dynamics \cite{Rivas10}. Note that choosing the environment as a collection of damped bosonic modes, one can use the Lindblad master equation in Eq.~\eqref{eq:lind} to describe the global dynamics, while the reduced dynamics of the electronic system $\rho_S(t)$ presents features of non-Markovianity. In what follows we analyze the dynamics under RWA, where the interaction Hamiltonian $H_{I}$ is energy conserving and aligned with the standard framework of thermal operations. 

\section{Generic dynamics}
\label{sec:suboptimal}

Motivated by the situations frequently encountered in biomolecular complexes where external illumination is either weak and/or doubly excited states are strongly suppressed, our analysis is restricted to the single excitation manifold. In this regime we monitor how a state initially populating only one electronically excited state, here $|e_1\rangle$, evolves due to the coupling to the vibrational environment. In the following, we focus on the finite-time dynamics generated by the model. We consider coupling strengths $g_i \ll \min_k \omega_k$ in ranges which correspond to population transfer timescales in line with experimentally observed reaction times for ultrafast photoisomerization \cite{Hahn}. Throughout this work, all energy gaps and mode frequencies are chosen to be of the order of the thermal energy $k_B T$. This choice ensures that the observed population transfer is genuinely thermodynamic, rather than being trivially dominated by either very large or very small energy gaps. In what follows we restrict ourselves to values where a considerable difference between the thermal yields Eq.~\eqref{eq:opt_yield} and Eq.~\eqref{eq:mark_yield} is observed. 

\begin{figure}[h]
    \centering
    \includegraphics[width=1\linewidth]{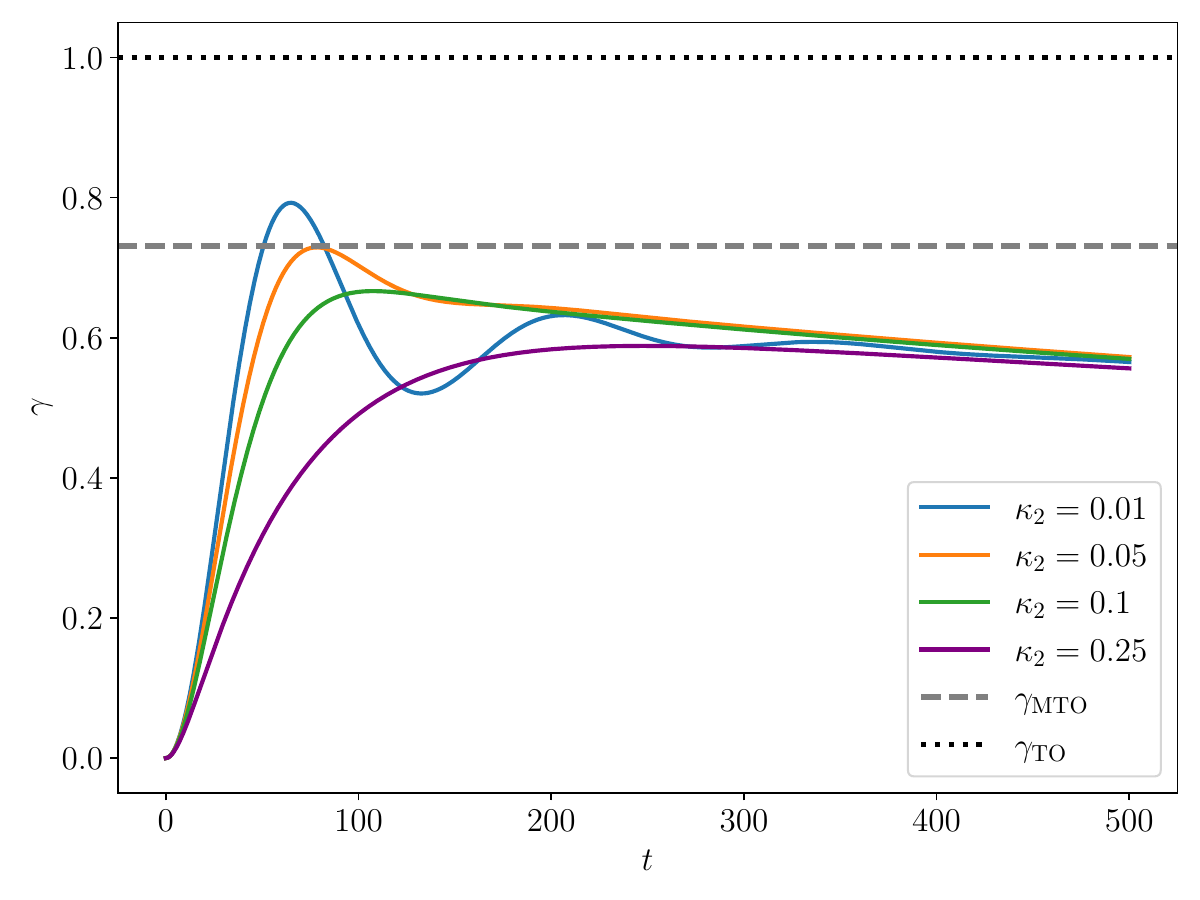}
    \caption{Time evolution of the yield $\gamma$ for different values of non-Markovianity controlled by damping rate of mode $2$. The molecule is initially fully excited $q = 1$, and its energy values are fixed at $\beta E_0 = 0$, $\beta E_1 = 2$, $ \beta E_2 = 1$, $ \beta E_3 = 1.5$. The three modes are truncated to $14, 12$ and $10$ Fock levels, respectively. The coupling and damping parameters are fixed at $g_1 = 0.01, g_2 = 0.02, g_3 = 0.01$ and $\kappa_1 = \kappa_3 = 0.4$. The horizontal lines correspond to the optimal thermal bound (dotted black) and the optimal Markovian thermal bound (dashed gray).
    }
    \label{fig:MY_nonopt}
\end{figure}

Figure~\ref{fig:MY_nonopt} shows the corresponding time evolution of the yield for fixed couplings $g_2 = 0.02$, $g_1 = g_3 = 0.01$ and damping rates $\kappa_1 = \kappa_3 = 0.4$. While strongly damped pseudomodes lead to dynamics fully compatible with the Markovian thermal bound, reducing the damping rate induces memory effects that enhance population transfer and eventually allow the yield to surpass the MTO bound. This behavior can be understood by noting that the electronic-vibrational coupling induces an effective relaxation rate for the electronic degrees of freedom. Whether the reduced electronic dynamics is Markovian or not is determined primarily by the vibrational damping we keep free $\kappa_2$, relative to the intrinsic timescales. Large values of $\kappa_2$ lead to rapid loss of excitations into the Markovian bath. In contrast, in the regime of small $\kappa_2$, memory effects become relevant and lead to a non-Markovian behavior, since the excitations take longer to fade from the mode, which benefits the population transfer, i.e. excitations travel more throughout the system before getting damped to the Markovian bath. Importantly, in this scenario the Markovian dynamics itself does not saturate the MTO bound. Even in the limit of strong damping, the yield remains strictly below $\gamma_{\mathrm{MTO}}$, indicating that the underlying Markovian evolution is thermodynamically sub-optimal. The non-Markovianity degree $\overline{\mathcal{N}}_{\rm RHP}$ can be computed by employing the Rivas-Huelga-Plenio (RHP) measure \cite{Rivas10} which evaluates the divisibility of the dynamical map over the chosen parameter range and fixed time interval (see Appendix \ref{ap:NM_quantification} for details). A nonzero value of the measure indicates that the reduced dynamics is not CP-divisible. When a canonical time-local generator exists, this loss of CP divisibility is associated with at least one temporarily negative decay rate. It does not, in general, preclude the existence of a time-local master equation. 
When the dynamical map is CP-divisible, $\overline{\mathcal{N}}_{\rm RHP}$ vanishes and the evolution is fully Markovian. For the parameter set considered in Fig.~\ref{fig:MY_nonopt}, we obtain for $\kappa_2 = 0.01$, $\kappa_2 = 0.05$, $\kappa_2 = 0.1$ and $\kappa_2 = 0.25$, yielding $\overline{\mathcal{N}}_{\rm RHP} = 0.087$, $\overline{\mathcal{N}}_{\rm RHP} = 0.071$, $\overline{\mathcal{N}}_{\rm RHP} = 0.031$, and $\overline{\mathcal{N}}_{\rm RHP} = 0.0003$, respectively.

We find that this behavior is robust across the parameter space: an enhancement in the transient dynamics occurs in the non-Markovian regime when compared to the Markovian one. However, the mere presence of non-Markovianity does not guarantee optimal thermodynamic performance. This highlights a central message of our work: although memory effects are \textit{necessary} to surpass Markovian thermal bounds, the ones induced by the present model, are not sufficient to reach the optimal efficiency allowed by TO. Only when non-Markovianity is carefully engineered so as to reinforce the population transfer toward the target level does it provide a significant finite-time thermodynamic advantage. This is the focus of the next section. 

\section{Optimally structured non-Markovian dynamics}
\label{sec:optimal}

We now turn to a class of dynamics in which memory effects are deliberately engineered. Before moving to the investigation of the dynamics considered in this section, it is important to clarify the role of the thermal bounds and their relation to dynamical optimality. The bounds are obtained by optimizing over the full class of thermal operations and are therefore tight, characterizing the maximal performance achievable under this operational restriction. However, this optimization does not imply that arbitrary thermal dynamics, Markovian or not, will saturate or exceed these bounds. On the contrary, as shown in Sec.~\ref{sec:suboptimal}, most admissible dynamics remain sub-optimal, since population can be redistributed through several competing transitions. The central question is therefore not whether non-Markovianity is present, but how it is structured to improve thermodynamic efficiency.

\begin{figure}[t]
    \centering
    \includegraphics[width=1\linewidth]{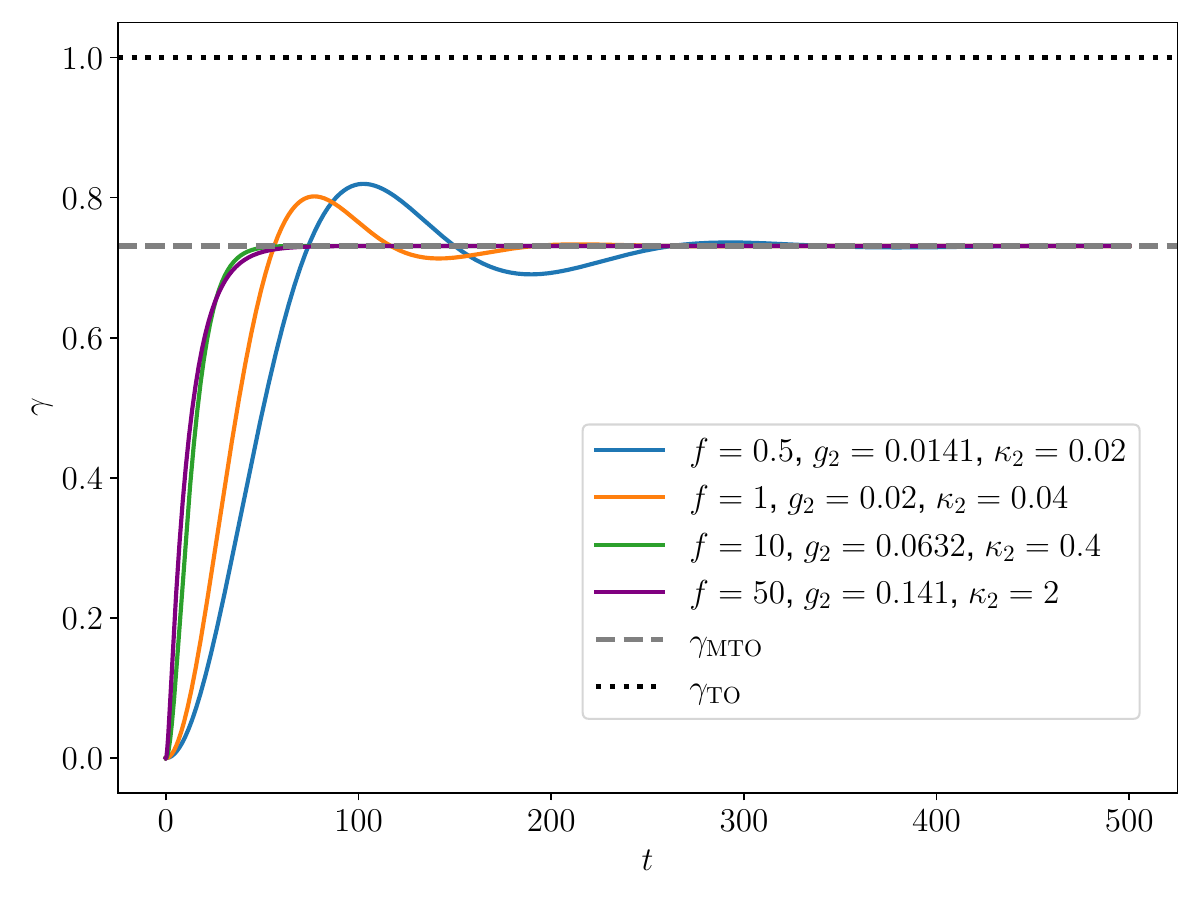}\\
    \includegraphics[width=1\linewidth]{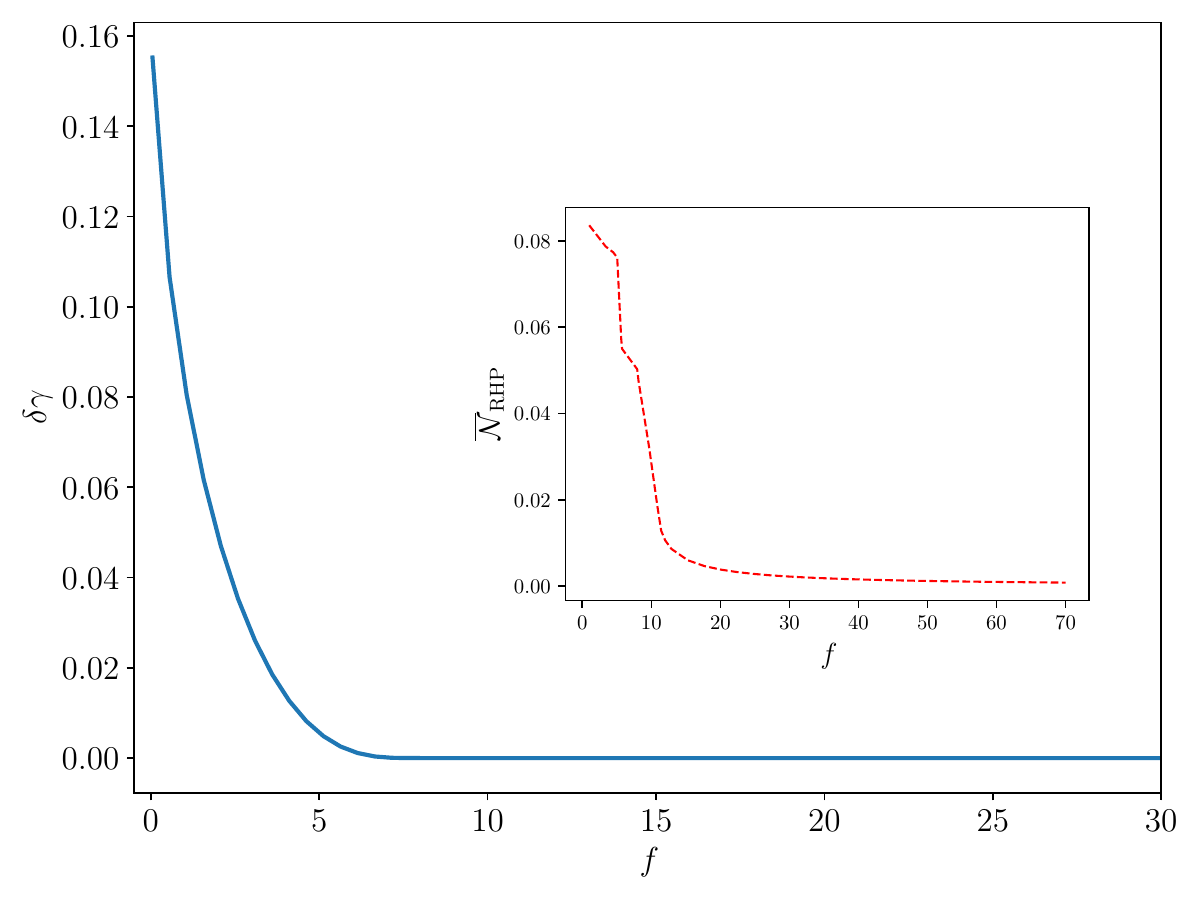}
    \caption{Top: Time evolution of the yield $\gamma$ as a function of non-Markovianity controlled by the parameter $f$. Energy values are fixed at $\beta E_0 = 0$, $\beta E_1 = 2$, $\beta E_2 = 1$, $\beta E_3 = 1.5$, and the molecule is initially fully excited $q = 1$. The pseudomode is truncated to twelve levels, $g_0= 0.02, \kappa_0 = 0.04$. Bottom panel: Relative yield advantage $\delta \gamma = (\gamma_{\rm max} - \gamma_{\rm MTO})/\gamma_{\rm MTO}$ as a function of $f$. In the inset we depict the degree of non-Markovianity $\overline{\mathcal{N}}_{RHP}$ as a function of $f$ (see Appendix~\ref{ap:NM_quantification} for details). The dynamics is tracked over the time interval $I = [0,500]$, which fully encompasses the relevant memory features of the system. Numerical sampling was performed with a step size of $\Delta t = 10^{-2}$ which was chosen to ensure numerical convergence (further decreasing $\Delta t$ yields identical results). Decreasing $f$ increases non-Markovianity and the dynamics eventually achieves a yield that exceeds $\gamma_{\mathrm{MTO}}$ when non-Markovianity becomes non-vanishing for sufficiently small $f$ thus demonstrating that non-Markovianity can act as a genuine resource for improving energy transfer efficiency.} 
    \label{fig:MY_RWA}
\end{figure}

To this end, we performed a numerical search over the coupling strengths $\mathbf{g} = (g_1, g_2, g_3)$ and damping rates $\boldsymbol{\kappa} = (\kappa_1, \kappa_2, \kappa_3)$ to obtain the largest sampled yield 
\begin{equation}
    \gamma_{\rm{opt}}= \max_{(\mathbf{g},\boldsymbol{\kappa})\in\Lambda}\max_{t\in[0, t_{\max}]}\left\langle e_2 \middle| \rho_S(t) \middle| e_2 \right\rangle\,.
\end{equation}
Within the searched parameter domain $\Lambda = \{(\mathbf{g}, \boldsymbol{\kappa})|g_i \in [0, 0.2], \kappa_i\in[0.02, 4.0],\,i=1,2,3\}$, and for the fixed initial state, energies and time interval $t_{\max}$, the largest yield occurred on the boundary
\begin{equation}
g_1=g_3=0.
\end{equation}
The parameter ranges were chosen so that the reduced electronic dynamics could explore both Markovian and non-Markovian regimes for each pseudomode. This numerical result suggests that, within the investigated domain, the target yield is favored by suppressing couplings on the competing transitions while retaining the structured mode on the target transition $|e_1\rangle\leftrightarrow|e_2\rangle$. We therefore adopt this reduced architecture in the remainder of the section.

Observe that for the dissipator in Eq.~\eqref{eq:diss}, the relevant correlation functions of the pseudomode are
\begin{align}
\langle a_2(t)a_2^\dagger(0)\rangle&=(\bar n_2+1)e^{-(\kappa_2/2+i\omega_2)t},\\
\langle a_2^\dagger(t)a_2(0)\rangle&=\bar n_2e^{-(\kappa_2/2-i\omega_2)t}.
\end{align}
The environmental correlations decay exponentially on a timescale determined by the inverse of the damping rate, so that 
\begin{equation}
\label{eq:env_time}
\tau_E\propto\frac{1}{\kappa_2}\,. 
\end{equation}
When the pseudomode relaxes rapidly compared with the system-mode excitation exchange, its effect on the electronic system can be described by an effective relaxation rate that scales as $\sim g_2^2/\kappa_2$ \cite{Lemmer,Reiter2012}. An enhancement observed upon decreasing $\kappa_2$ could therefore result from the longer environmental memory, from the accompanying increase in the effective transition rate, or from a combination of both effects. To separate these contributions, we introduce the one-parameter family  
\begin{equation}
\label{eq:f}
    g_2(f) = \sqrt{f}\, g_0\,, \quad \kappa_2(f) = f\, \kappa_0\,,
\end{equation}
for constants $g_0$ and $\kappa_0$. Along this family $g^2_2(f)/\kappa_2(f) = g_0^2/\kappa_0$ is constant. Consequently, in the weak-coupling, rapidly damped regime, the leading-order effective transition rates, which scale as $g_2^2/\kappa_2$, are held fixed~\cite{Reiter2012}. Higher-order corrections and the complete microscopic dynamics need not remain unchanged. At the same time, the environmental correlation time decreases with increasing $f$. This parametrization therefore varies finite-memory effects while preserving the leading-order effective dissipation scale. Thus, increasing $f$ suppresses finite-memory effects and drives the dynamics towards the Markovian limit without changing the leading transition rates. We emphasize that this parametrization does not leave the complete microscopic dynamics unchanged, since $g_2$ and $\kappa_2$ vary individually. Rather, it provides a controlled comparison in which finite correlation time corrections are varied while the leading effective dissipation rate is held fixed \cite{NM_assisted}.

We set $g_0=0.02, \kappa_0=0.04$, and let $f\in[0.1,100]$. In this case, $g_2^2/\kappa_2=0.01$. In Fig. ~\ref{fig:MY_RWA} (top), we depict the population dynamics for fixed energy values, and compare our results with the corresponding resource-theoretic bounds obtained for TO and MTO respectively (black dotted and gray dashed lines, respectively). The molecule is initially fully excited, $q =1$. The mode is truncated to twelve levels, since the probability of occupying higher Fock states is negligible in the considered parameter regime. The lower panel shows the advantage provided by non-Markovianity in overcoming the Markovian thermal bound. The advantage is quantified by comparing the maximal yield $\gamma_{\max}$ achieved within the time interval of interest to the corresponding Markovian thermal bound. In particular, the relative advantage $\delta \gamma$ assesses how much $\gamma_{\max}$ exceeds this benchmark, thereby capturing the relative enhancement over $\gamma_{\rm MTO}$. We find that the resource-theoretic yield for Markovian bounds is exceeded only when $\overline{\mathcal{N}}_{\rm RHP}>0$ (see the inset of the lower panel of Fig.~\ref{fig:MY_RWA}). For smaller $f$, the reduced dynamics is more non-Markovian and the maximal yield exceeds the Markovian thermal bound. As $f$ increases, the RHP indicator approaches zero and the advantage disappears. The observed enhancement therefore persists when the leading effective relaxation scale is held fixed and cannot be attributed solely to the accompanying change in $g_2^2/\kappa_2$.

A key observation emerging from Fig.~\ref{fig:MY_RWA} is that non-Markovianity is not merely correlated with enhanced performance, but constitutes a \textit{necessary} condition for surpassing the MTO bound. For all parameter regimes in which the reduced dynamics remains Markovian, $\overline{\mathcal{N}}_{\mathrm{RHP}} = 0$, the yield never exceeds $\gamma_{\mathrm{MTO}}$. Conversely, violations of the Markovian thermal bound are observed exclusively when the dynamics becomes non-Markovian, as witnessed by a strictly positive RHP measure.

\begin{figure}[h!]
    \centering
    \includegraphics[width=1\linewidth]{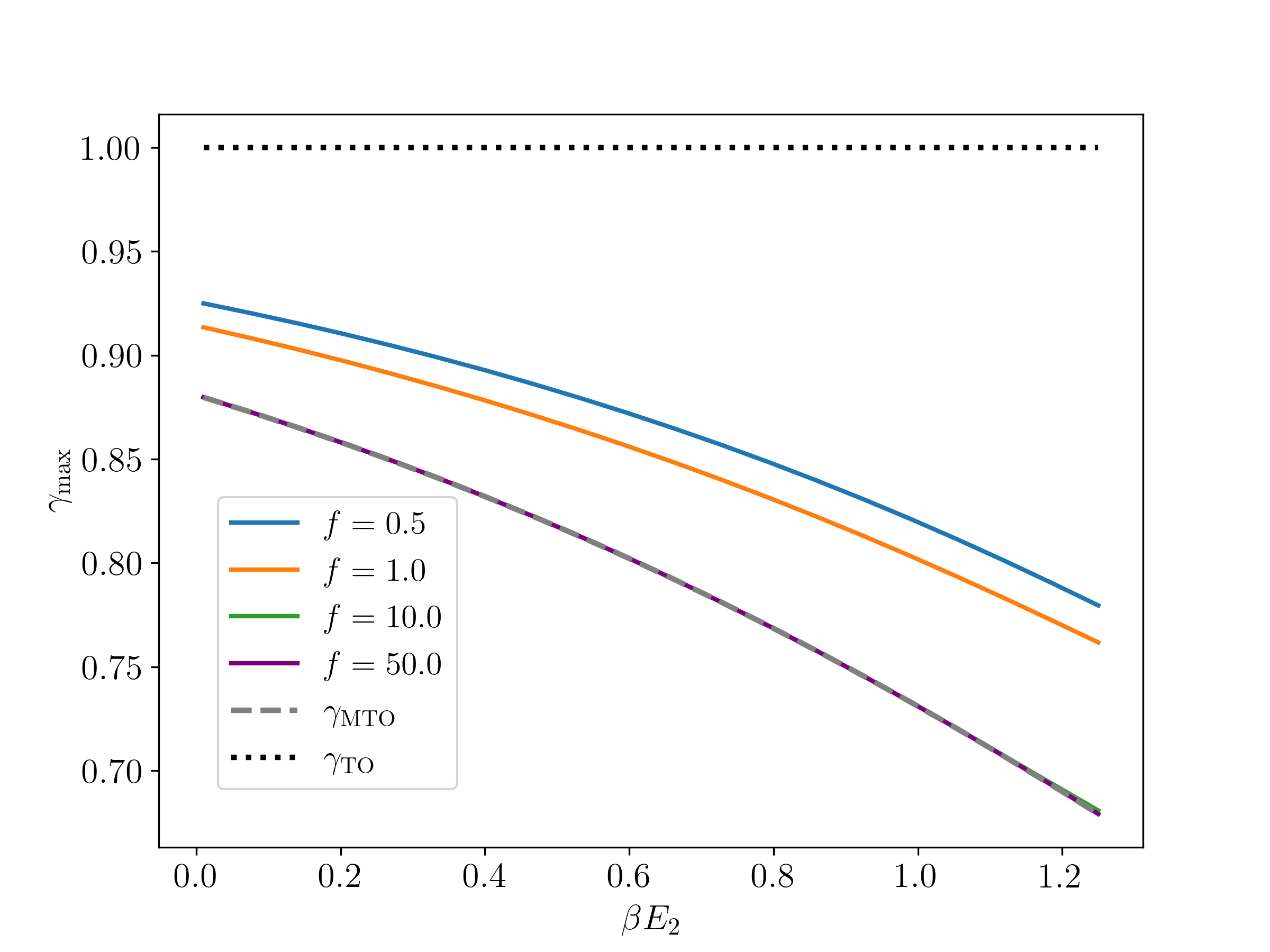}
    \caption{Maximum yield $\gamma_{\rm max}$ as a function of the energy of the target level $\beta E_2$ for different $f$-values. Energy values are fixed at $\beta E_0 = 0$, $\beta E_1 = 2$, $\beta E_3 = 1.5$, and the molecule is initially fully excited $q = 1$. The mode is truncated to twelve levels, with $g_0= 0.02$ and $\kappa_0 = 0.04$. For each energy value, we tracked the time evolution over the interval $I = [0,500]$. For $f = 10$ and $f=50$ the curves coincide with the Markovian thermal bound.} 
    \label{fig:MaxY}
\end{figure}

Further insight into the structured nature of the dynamics is provided by Fig.~\ref{fig:MaxY}, where we plot the maximum achievable yield as a function of the target energy level $\beta E_2$. The dependence of $\gamma_{\mathrm{max}}$ on $E_2$ closely follows the thermodynamic expectations dictated by the energy landscape of the molecule. The model responds smoothly to changes in the energetic configuration, reinforcing the conclusion that the observed non-Markovian advantage reflects a genuine thermodynamic mechanism rather than fine-tuned control.

The optimization here identifies the architecture that maximizes the population transferred to the target level: the target transition remains coupled to a structured vibrational mode, whereas the competing transitions are suppressed. We find the same qualitative behavior when the pseudomodes coupled to these competing transitions are replaced by effective thermal Lindblad dissipators, as expected in their rapidly relaxing, Markovian limit. In the following section, we offer an illustrative discussion on the broader set of population vectors accessible to the dynamics and their inherent limitations.

\section{Reachable set of states}
\label{sec:cones}

While the yield of the process provides a direct measure of efficiency, resource theories also characterize entire sets of possible state transformations. To assess to which degree realistic dynamics approximately exhaust the potential and limits imposed by resource theories, we analyze the regions of the state space accessible under our model and compare them with the regions accessible by thermal operations. Within the resource theory of thermodynamics, the set of reachable states from an initial state under thermal operations is often referred to as \textit{thermal cones}
\begin{equation}
    \mathcal{C}_{\mathrm{TO}}(\rho_S)=\left\{\sigma_S\middle|\,\rho_S\succ_{\mathrm{th}}\sigma_S\right\}\,, 
\end{equation}
where $\succ_{\mathrm{th}}$ denotes thermomajorization \cite{LostaglioReview}.
For a state that is diagonal in the energy eigenbasis, the set of all physical population vectors is the three-dimensional probability simplex \cite{Boyd2004}
\begin{equation}
    \Delta_3=\left\{ \mathbf{p}\in\mathbb{R}^{4}_{+}\,\middle|\,\sum_{i=0}^{3}p_i=1\right\}. 
\end{equation}
For the family of initial states considered here, we denote the initial population vector as $\mathbf{p}_0$, see Eq.~\eqref{eq:rhoSini}. The initial excited state population is denoted by the photoexcitation factor $q$.
\begin{equation}
    \mathcal{C}_{\mathrm{TO}}(q)=\left\{\mathbf{p}\in\Delta_3\,\middle|\,\mathbf{p}_0\succ_{\mathrm{th}}\mathbf{p}\right\},
\end{equation}
This polytope can be constructed from the extremal states associated with the corresponding $\beta$-permutations \cite{LostaglioETO}.  

We use the reduced dynamic model identified in the previous section: the target transition is coupled to one damped vibrational mode, while the remaining electronic transitions are described by incoherent Lindblad rates $\Gamma_i$. The Lindblad generator is then given by 
\begin{align}
\label{eq:lindNew}
\mathcal{L}_\theta(\rho) &= -i [H_S + \omega_2 a_2^\dagger a_2 + g_2(A_2a_2^\dagger + A_2^\dagger a_2), \rho ] \nonumber\\
&\quad+ \kappa_2 (1+ \bar{n_2} )D[a_2](\rho) + \kappa_2 \bar{n}_2 D[a_2^\dagger](\rho)\nonumber\\
&\quad+ \sum_{i \in \{1,3\}} \Gamma_i \left[ (1+ \bar{n_i} ) D[A_i](\rho) + \bar{n}_i D[A_i^\dagger] (\rho)]\right]\,,
\end{align}
where $D[L](\rho)$ denotes the dissipative contribution as in Eq.~\eqref{eq:diss2}. In this case, the pseudomodes associated with the competing transitions relax rapidly and their reduced action becomes effectively Markovian. In contrast, the pseudomode coupled to the target transition is allowed to access non-Markovian regimes.

\begin{figure*}[t!]
    \centering
    \includegraphics[width=0.8\linewidth]
    {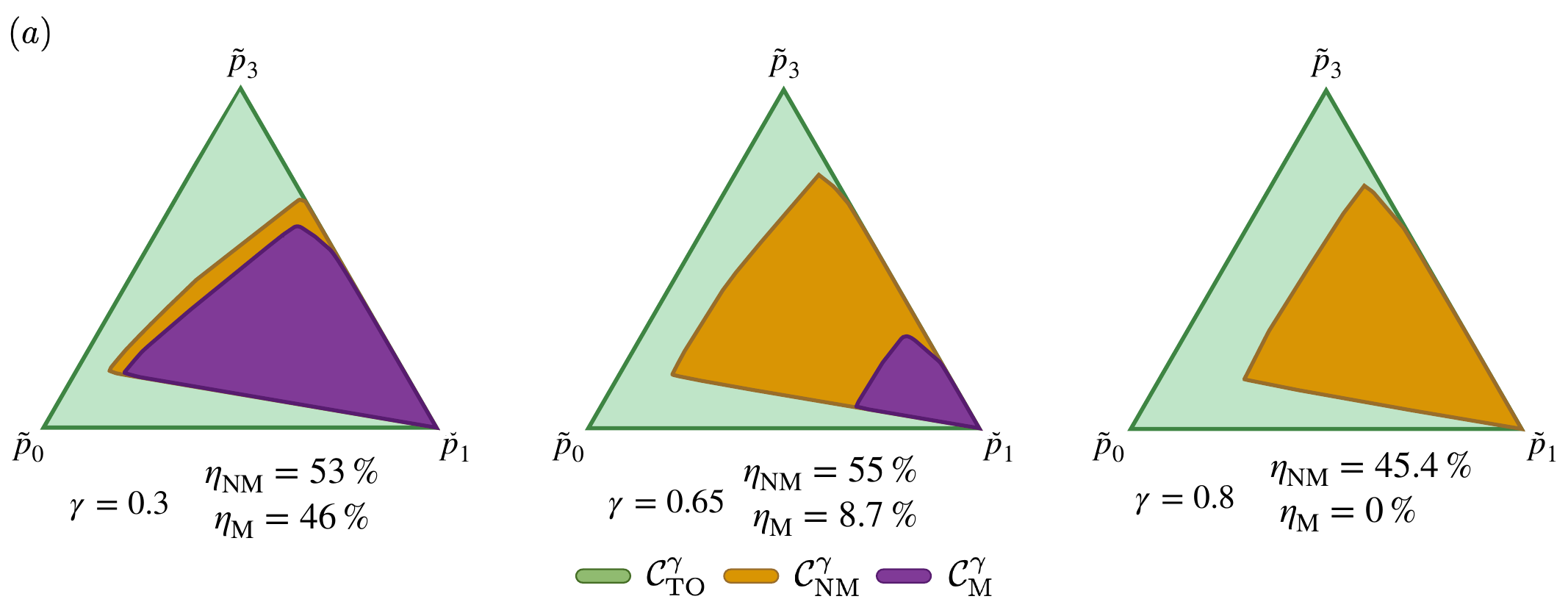}\\
    \includegraphics[width=0.8\linewidth]
    {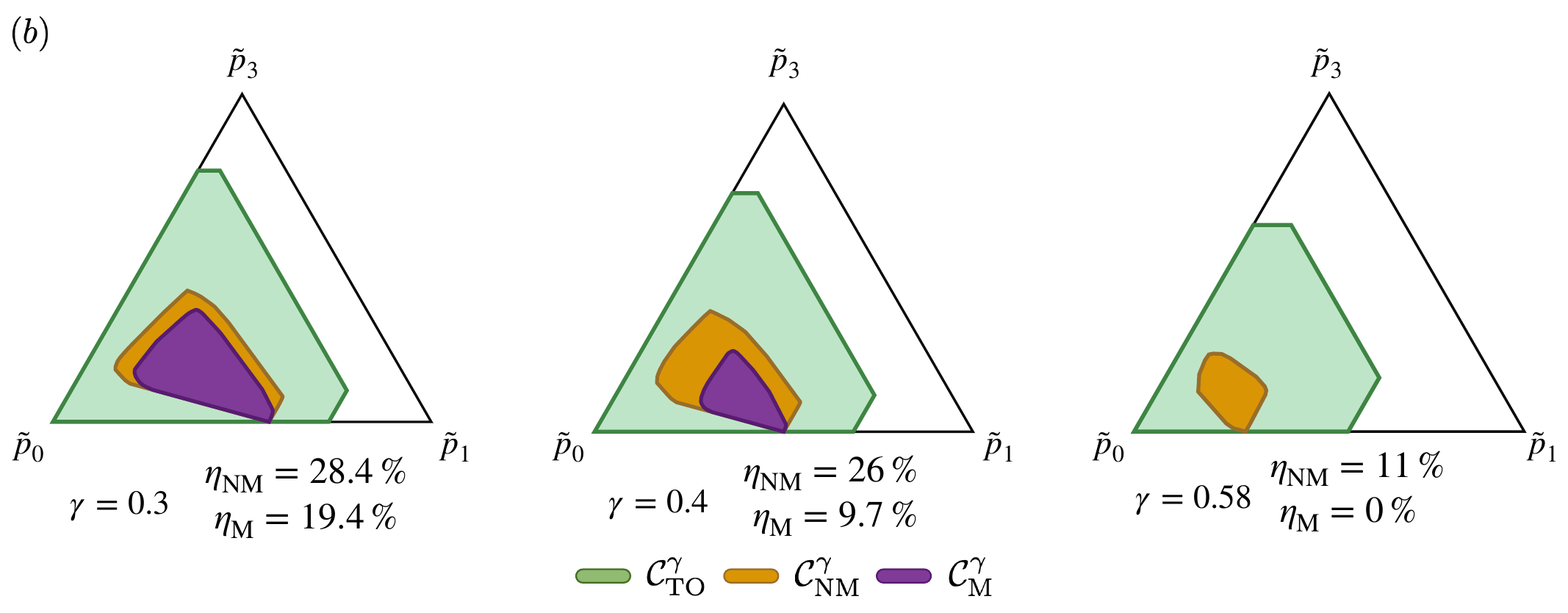}
    \caption{Enlarged dynamical accessibility facilitated by thermal non-Markovian operations. Depicted are the cross-sections of the sampled reachable sets at fixed target population $\gamma$, for the initially excited state $\mathbf{p}_0(q)=(1-q,q,0,0)^{\mathsf T}$ and energies $\beta\mathbf{E}=(0,2,1,1.5)$. 
    Each triangle represents the normalized
    residual populations $(\tilde p_0,\tilde p_1,\tilde p_3)$. The plotted polygons are the intersections of the Markovian and non-Markovian model convex hulls with the plane $p_2 = \gamma$. The displayed percentages are the TO coverages $100\,\eta_{\rm M}$ and $100\,\eta_{\rm NM}$. The dynamical hulls are constructed from a sample over $g_2\in[0.01,0.1], \kappa_2\in[0.02,2.5]$, and $\Gamma_1,\Gamma_{3}\in[10^{-4},0.1]$. The sampling comprises an initial global design of $6^4$ points, and one adaptive refinement round adding $64$ parameter points. For each sampled parameter tuple, populations are collected over the evolution interval $t\in[0,8000]$ and used to construct the corresponding reachable-state hull. (a) Fixed $q = 1$, the TO section coincides with the full outer triangle. The panels show the representative values $\gamma=0.3, 0.65$, and $0.8$, spanning target populations below and above the Markovian thermal bound $\gamma_{\rm MTO} = 0.73$. (b) Fixed $q=0.7$. The panels show the representative values $\gamma=0.3, 0.4$, and $0.58$, spanning target populations below and above the Markovian thermal bound $\gamma_{\rm MTO} = 0.57$.}
    \label{fig:cones}
\end{figure*}

Let $\boldsymbol{\theta} = (g_2, \kappa_2, \Gamma_1, \Gamma_3)$ collect the coupling and damping parameters of this model. For each choice of parameters, the reduced evolution defines the family of channels $\{\mathcal{E}_t^{\boldsymbol{\theta}}\}_{t \in [0, t_{\max}]}$ with 
\begin{equation}
    \mathcal{E}^{\boldsymbol{\theta}}_t(\rho_S)=
    {\rm Tr}_{E}\left[
        e^{t\mathcal{L}_{\boldsymbol{\theta}}}
        \left(\rho_S\otimes\rho_E\right)\right], 
\end{equation}
where $\mathcal{L}_{\boldsymbol{\theta}}$ is the Lindblad generator associated with the parameters in $\boldsymbol{\theta}$. The Markovianity classification is assigned to the complete process family of maps $\{\mathcal{E}_t^{\boldsymbol{\theta}}\}_{t \in [0, t_{\max}]}$, rather than to an individual map at a single time. In particular, using the RHP criterion, we divide the sampled operations into
\begin{align*}
    O_{\rm M} &= \big\{\mathcal{E}_t^{\boldsymbol{\theta}} : \overline{\mathcal{N}}_{\rm RHP} = 0\big\}\,,\\
    O_{\rm NM} &= \big\{\mathcal{E}_t^{\boldsymbol{\theta}}: \overline{\mathcal{N}}_{\rm RHP} \neq 0 \big\}\, .
\end{align*}
The previous sets will define a set of reachable states
\begin{align}
    \mathcal{S}_{X}(\rho_S) &= \Big\{ \sigma_S \mid \rho_S \xrightarrow[]{O_{X}} \sigma_S \Big\}\,,
    % \mathcal{S}_{\mathrm{NM}}(\rho_S) &= \Big\{ \sigma_S \mid \rho_S \xrightarrow[]{O_{\mathrm{NM}}} \sigma_S \Big\} \, , 
\end{align}
with $X\in\{\mathrm{M},\rm NM\}$. These sets are obtained by a uniform sample constructed by selecting a finite set of values for each component of $\boldsymbol{\theta}$ within the prescribed physical range. Thus, all states sampled along a given trajectory inherit the classification of the process that generates that trajectory. For every parameter tuple, we evolve the system on a uniform time grid and record
\begin{equation}
    p_i(t_n,\boldsymbol{\theta},q)=\langle i|
    \mathcal{E}^{\boldsymbol{\theta}}_{t_n}
    \bigl(\rho_S(q)\bigr)|i\rangle.
\end{equation}
The parameter ranges, number of grid points, evolution time, and temporal resolution are kept fixed when comparing the two dynamical classes. The sampling is sufficiently dense such that additional sampling points do not alter the shape or extent of the observed regions. In this sense, we approximate the boundary of the reachable region by the convex hull of the sampled states, used here as a numerical outer envelope. The hull closely matches the observed distribution of points and is a reliable estimate of the boundary of the accessible region in our parameter regime. We emphasize that this does not imply that the reachable set of our dynamics is itself convex. In general, the dynamics generated by a Lindblad master equation is not closed under convex mixing of maps, consequently, not every point inside the convex hull needs to be dynamically reachable. We use the convex hull of the sampled populations as an empirical approximation to the convexified reachable region within the investigated parameter domain. This approach ensures that all points are encompassed within the smallest convex shape that contains them $\mathcal{C}_{X}(q) = {\rm conv}\bigl(\mathcal{S}_{X}(q)\bigr)$. 

The full population simplex and the associated hulls are three-dimensional. To obtain a clearer visualization, we fix the target population
\begin{equation}
    p_2 = \gamma \qquad 0\leq \gamma<1,
\end{equation}
and consider the two-dimensional sections
\begin{equation}
    \mathcal{C}_{X}^{\gamma}(q)=
    \mathcal{C}_{X}(q)\cap\left\{\mathbf{p}\in\Delta_3\,\middle|\,p_2 = \gamma\right\},
\end{equation}
for $X\in\{\mathrm{M},\rm NM\}$. At fixed $\gamma$, the remaining populations obey $p_0+p_1+p_3=1-\gamma$. We therefore introduce the normalized residual populations
\begin{equation}
    \tilde{p}_j=\frac{p_j}{1-\gamma},\quad j\in\{0,1,3\},
\end{equation}
which satisfy $\tilde{p}_0+\tilde{p}_1+\tilde{p}_3=1$. They describe how the population not occupying the target level is distributed among the other three levels. The section polygon is obtained by intersecting the edges of the three-dimensional hull with the plane $\gamma$, followed by a two-dimensional convex-hull construction of the intersection points.

We embed this vector into an equilateral triangle through
\begin{equation}
    \Pi_\gamma(\mathbf{p})=\left(\tilde{p}_1+\frac{1}{2}\tilde{p}_3,\frac{\sqrt{3}}{2}\tilde{p}_3\right).
\end{equation}
The corresponding TO section is
\begin{equation}
    \mathcal{C}^\gamma_{\mathrm{TO}}(q)=
    \mathcal{C}_{\mathrm{TO}}(q)\cap\left\{\mathbf{p}\in\Delta_3\,\middle|\, p_2 = \gamma\right\}.
\end{equation}
In order to obtain a clear comparison between the sets, we define the area of the normalized section as
\begin{equation}
A_X(\gamma,q)=\operatorname{Area}\left[\Pi_\gamma\bigl(\mathcal{C}^\gamma_X(q)\bigr)\right].
\end{equation}
At fixed target population $\gamma$, the cross-sectional area $A_X(\gamma,q)$ quantifies the range of residual population distributions accessible under the considered class of dynamics. Several distinct dynamics may reach the same population vector, while a single trajectory may contribute several points to the section. A small value of the area means that the target population can be reached only together with a restricted set of distributions over the remaining levels. Conversely, a large area indicates greater flexibility in how the remaining population can be distributed while maintaining the same target value. Details of the construction of the reachable sets and the calculation of their areas are provided in Appendix~\ref{ap:cones}. 

We now define the TO coverage
\begin{equation}
    \eta_X (\gamma,q)=
    \frac{A_{\mathrm{X}}(\gamma,q)}
         {A_{\mathrm{TO}}(\gamma,q)}.
\end{equation}
Thus, $100\,\eta_X$ is the percentage of the TO-allowed section covered by the corresponding $X$ dynamical hull. 

In our simulations, the parameters are varied over $g_2 \in [10^{-2}, 10^{-1}]$, $\kappa_{2}\in[2\times10^{-2},2.5]$, $\Gamma_1\in[10^{-4},10^{-1}]$, and $\Gamma_{3}\in[10^{-4},10^{-1}]$. For each interval we compute 6 points, resulting in a global sample of $6^4$ points. For each sampled parameter tuple, populations are collected over the evolution interval $t\in[0,8000]$ and used to construct the corresponding reachable-state hull. To improve the resolution of the sampled reachable set, we perform an additional refinement round, in which 64 additional tuples are evaluated in regions identified as relevant to the hull boundary. 

Figure~\ref{fig:cones} first illustrates the construction for $q=1$. In this case, the system initially occupies the highest-energy level and $\mathcal{C}_{\mathrm{TO}}=\Delta_3$. Therefore, the figure isolates the difference between the Markovian and non-Markovian dynamical coverage. The enlargement of the cross-sectional area shows that the role of memory is not restricted to increasing the largest attainable value of $\gamma$. At a prescribed target population, it can also increase the range of compatible distributions over the remaining levels. A larger $\eta_{\rm NM}$ indicates that the non-Markovian model enables population transformations that are inaccessible under the certified Markovian dynamics, even though both classes reach the same value of $\gamma$. For partially excited initial states, $q<1$, the same construction yields a more stringent resource-theoretic comparison because the TO section is then a proper polygon inside the physical triangle. In panel (b) of Fig.~\ref{fig:cones}, we observe how the TO region decreases for increasing values of $\gamma$. For $\gamma > \gamma_{\rm MTO}$, no Markovian coverage is observed.

In principle, our model allows full and independent control over all system-mode couplings and damping rates. In practice, however, such flexibility is not available in realistic photoisomerization dynamics, where interaction strengths are constrained by the molecular structure and its surrounding environment. In biological settings in particular, these parameters are set by structural and energetic properties and cannot be varied arbitrarily. Accordingly, we restricted our analysis to a physically motivated parameter regime that reflects these inherent limitations. Accurate models of photoisomerization dynamics crucially depend on a faithful parametrization of the system-environment coupling consistent with the underlying potential energy landscape. Following photoexcitation, the system evolves along a well defined reaction coordinate, passing through a conical intersection that governs the isomerization pathway \cite{Limmer19, Nicole}. The present model is not intended to resolve the nuclear reaction coordinate or the conical-intersection dynamics explicitly. Instead, it provides an effective description of electronic population transfer in the presence of structured vibrational degrees of freedom, allowing the thermodynamic role of environmental memory to be isolated. Our results also highlight a resource-theoretic interpretation of non-Markovianity. Since stronger environmental memory effects consistently enlarge the region of dynamically accessible states, one may ask what states remain accessible when the degree of non-Markovianity is constrained below a fixed threshold. Investigating state reachability when imposing a memory budget on the RHP measure presents an interesting direction for future research. Our findings show that within the present model, environmental memory enlarges the dynamically accessible region and can enable population transfer efficiencies beyond the MTO bound.

\section{Conclusions}

In this work, we have investigated to what extent optimal performance bounds derived within thermodynamic resource theories can be approached by explicit microscopic dynamics. Thermal operations characterize state transformations that are possible under a prescribed set of thermodynamic constraints without specifying the microscopic environment or the dynamical mechanism implementing them. This generality is one of the main strengths of the resource-theoretic approach, but leaves open the question of whether transformations close to the resulting bounds can be realized within concrete and physically restricted system-environment architectures. Here, we have addressed this question using a tunable model in which an electronic system is coupled to a structured vibrational environment represented by damped pseudomodes.

Our results identify environmental memory as a mechanism that can lift dynamical restrictions associated with Markovian thermal evolutions, thereby enlarging the set of transformations accessible to the reduced system. This role is more general than the enhancement of a particular population-transfer yield. The reachable-state analysis shows that non-Markovian dynamics allows the microscopic model to explore a larger portion of the region permitted by thermal operations. Memory should therefore not be regarded simply as a resource whose presence or magnitude determines performance. Rather, it provides additional dynamical accessibility, allowing the reduced evolution to reach states that remain inaccessible under Markovian thermal dynamics. Enlarged accessibility alone, however, does not imply optimal performance. We showed that generic non-Markovian dynamics can remain substantially below the TO bound, since the additional dynamical freedom associated with memory need not favor the desired population-transfer pathway over competing ones. Approaching the resource-theoretic optimum depends on the microscopic structure of the system-environment coupling, which determines both the memory properties of the reduced dynamics and the pathways through which population is redistributed. Memory and coupling topology are therefore intrinsically intertwined: the relevant question is not how much memory is present, but whether the microscopic architecture generates dynamics that enlarge accessibility in directions favorable to the target transformation. In the present microscopic model for isomerization, suppressing competing transitions while retaining a structured environment along the productive pathway provides a concrete illustration of this interplay.

This perspective also clarifies the practical role of resource-theoretic bounds. Such bounds are not, by themselves, constructive prescriptions for microscopic dynamics, nor do we assume that naturally occurring chemical or biochemical processes are optimized to operate close to them. Rather, they provide model-independent benchmarks against which restrictions arising from a particular microscopic architecture can be distinguished from fundamental thermodynamic limitations. This distinction is particularly useful when the underlying microscopic description is incomplete. While predictions of the actual dynamics necessarily depend on the chosen model, the resource-theoretic bound remains independent of those details as long as its defining thermodynamic assumptions are satisfied. Conversely, confronting candidate microscopic models with the resource-theoretic bound can reveal how strongly their connectivity, coupling strengths, and environmental structure restrict transformations that remain thermodynamically possible in principle. Our findings suggest a natural route towards a more constructive use of thermodynamic resource theories. Rather than asking only whether a given microscopic model approaches a resource-theoretic optimum, one may invert the problem and ask which microscopic system-environment architectures are capable of realizing or approximating a prescribed transformation near the TO boundary. This would require characterizing the dynamical restrictions responsible for the gap between thermodynamic possibility and microscopic realizability, and determining which microscopic structures generate the additional dynamical accessibility required to overcome them.  Establishing such a connection between resource-theoretic reachability and the structure of microscopic dynamics could ultimately provide constructive principles for identifying environments capable of approaching fundamental performance bounds in specific tasks, from energy conversion and transport to state preparation and parameter estimation.

\section{Acknowledgments}
This work is financially supported by the DFG via QuantERA project ExTRaQT (Grant No. 499241080) and the European Research Council via the Synergy grant HyperQ (Grant no. 856432).

\bibliography{references}

%%________________________________________________
\begin{appendix}

\section{Engineering the interaction}
\label{ap:modes_thermal}

The thermal operations compose a large set of transformations, however they seem unrealistic from the experimental point of view due to the fact that in practical scenarios fundamental limitations in controlling the microscopic degrees of freedom of the bath emerge. In addition to being time-translation covariant, the interaction must be energy preserving. In particular, the global unitary implementing a thermal operation must commute with the total free Hamiltonian. For the present model, this condition reads
\begin{equation}
    \label{eq:comH}
    [H_I, H_S + H_E] = 0\,.
\end{equation}
To verify Eq.~\eqref{eq:comH}, we write
\begin{equation}
A_i=|e_{\beta_i}\rangle\langle e_{\alpha_i}|,\qquad\Delta_i:=E_{\alpha_i}-E_{\beta_i}>0,
\end{equation}
with
\begin{equation}
(\alpha_1,\beta_1)=(1,3),
\quad
(\alpha_2,\beta_2)=(1,2),
\quad
(\alpha_3,\beta_3)=(3,0).
\end{equation}
The electronic transition operators satisfy
\begin{equation}
[H_S,A_i]=-\Delta_i A_i,\qquad[H_S,A_i^\dagger]=\Delta_i A_i^\dagger,
\end{equation}
while the bosonic commutation relations imply
\begin{equation}
[H_E,a_i]=-\omega_i a_i,\qquad[H_E,a_i^\dagger]=\omega_i a_i^\dagger.
\end{equation}
It follows that
\begin{equation}
[H_S+H_E,A_i a_i^\dagger]=(\omega_i-\Delta_i)A_i a_i^\dagger,
\end{equation}
and 
\begin{equation}
[H_S+H_E,A_i^\dagger a_i]=(\Delta_i-\omega_i)A_i^\dagger a_i.
\end{equation}
Consequently,
\begin{equation}
[H_S+H_E,H_I]=\sum_{i=1}^{3}g_i(\omega_i-\Delta_i)\left(A_i a_i^\dagger-A_i^\dagger a_i\right).
\end{equation}
The commutator therefore vanishes when
\begin{equation}
\label{eq:resonance}
\omega_i=\Delta_i=E_{\alpha_i}-E_{\beta_i}.
\end{equation}
Thus, to satisfy the energy preserving condition, the frequencies of the modes must be resonant with the corresponding energy of the electronic transitions. 

It remains to verify that the damping of the pseudomodes preserves the TO structure, and that the evolution falls into the TO class. Let 
\begin{equation}
    \mathcal L_i^{\mathrm{th}}(X):=\kappa_i(1+\bar n_i)\mathcal D[a_i](X)+\kappa_i\bar n_i\mathcal D[a_i^\dagger](X)
\end{equation}
denote the thermal Lindblad generator acting on mode $i$ (Eq.~\eqref{eq:diss}), we omitted the identity on the other subsystems for simplicity. For any time step $\delta>0$, the channel
\begin{equation}
    \mathcal A_{i,\delta} := e^{\delta\mathcal L_i^{\mathrm{th}}}
\end{equation}
is a single-mode thermal attenuator with transmissivity $\eta_i=e^{-\kappa_i\delta}$ \cite{SerafiniBook}. To construct its dilation, introduce an ancillary bosonic mode $b_i$ of frequency $\omega_i$, initially prepared in its Gibbs state
\begin{equation}
    \tau_{b_i}=\frac{e^{-\beta\omega_i b_i^\dagger b_i}}{{\rm Tr}\left[e^{-\beta\omega_i b_i^\dagger b_i}\right]}.
\end{equation}
The attenuator admits the dilation
\begin{equation}
    \mathcal A_{i,\delta}(X)={\rm Tr}_{b_i}\left[U_{i,\delta}(X\otimes\tau_{b_i})U_{i,\delta}^\dagger\right],
\end{equation}
where
\begin{equation}
    U_{i,\delta}=\exp\!\left[ \theta_{i,\delta}\left(a_i^\dagger b_i-a_i b_i^\dagger\right)\right],
\end{equation}
with $\cos^2\theta_{i,\delta}=e^{-\kappa_i\delta}$. This unitary strictly conserves the total bare energy,
\begin{equation}
    \left[U_{i,\delta}\,,\,\omega_i\left(a_i^\dagger a_i+b_i^\dagger b_i\right)\right]=0.
\end{equation}
Consequently, we have shown that the damping of the pseudomodes does not erase energy, it exchanges excitations from the pseudomode to a thermal reservoir oscillator. 
The coherent evolution over the same time step is
\begin{equation}
    \mathcal U_\delta(X)=W_\delta XW_\delta^\dagger,
\end{equation}
with $W_\delta=e^{-i(H_S+H_E+H_I)\delta}$. From the resonance condition, established in Eq.~\eqref{eq:resonance}, we get 
\begin{equation}
    [W_\delta,H_S+H_E]=0,
\end{equation}
so this step is also energy preserving. The global generator is then 
\begin{equation}
    \mathcal{L}(\cdot)=-i[H_S+H_E+H_I,\,\cdot\,]+\sum_{i=1}^3\mathcal L_i^{\mathrm{th}}(\cdot)\,.
\end{equation}
The complete evolution follows from the Lie-Trotter formula,
\begin{equation}
    e^{t\mathcal L}=\lim_{n\rightarrow\infty}\left[\mathcal A_{3,t/n}\circ\mathcal A_{2,t/n}\circ\mathcal A_{1,t/n}\circ\mathcal U_{t/n}\right]^n.
\end{equation}
At every finite $n$, the dilation may be implemented using fresh ancillary modes prepared in Gibbs states at inverse temperature $\beta$. Every constituent unitary commutes with the corresponding total free Hamiltonian, and hence so does their product. Since the pseudomodes are initially prepared in the Gibbs state, tracing out both the pseudomodes and the ancillary reservoir modes therefore produces a thermal operation on the electronic system.

Thus, for every $t$, the reduced electronic map
\begin{equation}
    \mathcal E_t(\rho_S)={\rm Tr}_E\left[e^{t\mathcal L}(\rho_S\otimes\tau_E)\right]
\end{equation}
is obtained as the continuous-time limit of thermal operations. Equivalently, it belongs to the closure of the TO class, which is sufficient here because the thermomajorization region is closed \cite{LostaglioReview, vomEnde22}. In particular,
\begin{equation}
\langle e_2|\mathcal E_t(\rho_S)|e_2\rangle\leq\gamma_{\rm TO}
\end{equation}
for every $t$.

%%________________________________________________
\section{Quantification of the degree of non-Markovianity of the dynamics}
\label{ap:NM_quantification}

In this section we provide an explicit calculation of the RHP measure of non-Markovianity, first proposed in \cite{Rivas10} and widely explored in the pseudomode formalism \cite{NM_assisted, Lemmer}. Let us consider an open quantum system whose time evolution is described by a CPTP map $\mathcal{E}_{t, t_0}$, such that 
\begin{equation}
 \rho(t) = \mathcal{E}_{t, t_0}\rho(t_0).
\end{equation}
The map can be expressed as a composition of maps 
\begin{equation}
 \mathcal{E}_{t_2, t_0} = \mathcal{E}_{t_2, t_1} \mathcal{E}_{t_1, t_0}. 
\end{equation}
We say that the evolution is Markovian if and only if the map $\mathcal{E}_{t_2, t_1}$ exists and is CPTP for all $t_2 > t_1 > t_0$. The RHP measure $\mathcal{N}_{\mathrm{RHP}}$ then quantifies the total departure from complete positivity of these intermediate maps over a time interval $I$, defined as 
\begin{equation}
\label{eq:NRHP}
\mathcal{N}_{\mathrm{RHP}} = \int_{I} h(t) \,dt,
\end{equation}
with 
\begin{equation}
\label{eq:gt}
h(t) = \lim_{\epsilon \to 0^+}\frac{\parallel \chi(t+\epsilon, t) \parallel_1 -1}{\epsilon},
\end{equation}
where $\chi(t+\epsilon, t) = [\mathcal{E}_{t + \epsilon,t}\otimes \mathbb{1}]|\Phi^+\rangle\langle\Phi^+|$ corresponds to the Choi matrix of the intermediate map, such that $|\Phi^+\rangle = \sum_{k=0}^{d-1} |k\rangle|k\rangle/\sqrt{d}$ is the maximally entangled state for a system of dimension $d$. The quantity $h(t)$ vanishes whenever the intermediate map $\mathcal{E}_{t+\epsilon,t}$ is completely positive. More precisely, for a trace-preserving intermediate map,
\begin{equation}
    \left\|\chi(t+\epsilon,t)\right\|_1 \geq 1,
\end{equation}
with equality if and only if the Choi matrix is positive semidefinite. Consequently, $h(t)>0$ signals a violation of complete positivity and hence a breakdown of CP divisibility at time $t$.
The numerical computation of the measure requires the evaluation of a discrete version of Eq.~\eqref{eq:NRHP}. To evaluate the dynamical map, we divide the time interval $ I = [0, t_{\max}] $ into $N$ equally spaced discrete time steps $t_i $, with $t_0 = 0 $ and $t_N = t_{\max} $. We then compute the time evolution of the operator basis elements $|k\rangle\langle j| $, for $k, j = 0, 1, 2, 3 $, corresponding to a four-level system. For each such operator, the evolved state at time $t_i $ is denoted $|k\rangle\langle j|(t_i) = \rho_{kj}(t_i) $.

Each of these time-evolved operators can be \textit{vectorized} into a column vector:
\begin{align}
v_{kj}(t_i) = [&
\rho_{kj,00}(t_i),\, \rho_{kj,01}(t_i),\, \rho_{kj,02}(t_i),\,\rho_{kj,03}(t_i),\nonumber\\
&\rho_{kj,10}(t_i), \,\rho_{kj,11}(t_i), \rho_{kj,12}(t_i),\,\rho_{kj,13}(t_i),\nonumber\\
&\rho_{kj,20}(t_i),\rho_{kj,21}(t_i),\,\rho_{kj,22}(t_i),\,\rho_{kj,23}(t_i),\nonumber\\
&\rho_{kj,30}(t_i),\,\rho_{kj,31}(t_i),\,\rho_{kj,32}(t_i),\,\rho_{kj,33}(t_i)]^\mathsf{T}.
\end{align}
With these vectors, the dynamical map $\mathcal{E}_{t,t_0} $ can be represented in matrix form as
\begin{equation}
    V(t, t_0) = \big[v_{00}(t), v_{01}(t), \ldots, v_{33}(t) \big],
\end{equation}
where each column $v_{kj}(t) $ corresponds to the evolution of the basis operator $|k\rangle\langle j| $ at time $t $. The intermediate map can be constructed as 
\begin{equation}
V(t_{i+1},t_i) = V(t_{i+1},t_0)V(t_i,t_0)^{-1}\,,
\end{equation}
which assumes that $V(t_i,t_0)$ is invertible. In the numerical implementation, we use a direct linear solver whenever the smallest singular value of $V(t_i,t_0)$ is larger than $10^{-10}$ times its largest singular value. When this condition is not satisfied, we instead compute the pseudoinverse with relative singular-value cutoff $10^{-10}$.

Finally, the Choi matrix $\chi(t_{i+1}, t_i)$ is equal to the reshuffled matrix $\frac{1}{d}V^{\mathrm{R}}(t_{i+1},t_i)$. The reshuffling operation transforms the superoperator $V(t_{i+1},t_i)$, represented in the Liouville basis, into the Choi matrix $\chi(t_{i+1},t_i)$ on a $d$-dimensional Hilbert space \cite{Zycz}. For two consecutive times $t_i$ and $t_{i+1}$, with
$\Delta t=t_{i+1}-t_i$, we can then define an equivalent function $h(t)$ that computes the right derivative of the trace norm as 
\begin{equation}
 h(t_i) :=\frac{\left\|\frac{1}{d}V^{R}(t_{i+1},t_i)\right\|_1-1}{\Delta t},
\end{equation}
where $\|\cdot\|_1$ denotes the Schatten trace norm. Observe that $h(t)>0$ for some $t$ if and only if the evolution is non-Markovian. Since small negative values may arise from finite numerical precision, we retain only the positive part
\begin{equation}
    h^{+}(t_i):=\max\{0,h(t_i)\}.
\end{equation}
The original RHP integral is then approximated by
\begin{equation}
    \mathcal{N}_{\mathrm{RHP}}\simeq\sum_{i=0}^{N-1}h(t_i)^{+}\Delta t.
\end{equation}
For the comparison of different dynamical regimes, it is convenient to introduce a bounded (normalized) indicator. Following the bounded transformation employed in Ref.~\cite{Lemmer}, we map the instantaneous rate according to $h(t_i)^{+}\mapsto\tanh[h(t_i)^{+}]$. We thus
define
\begin{equation}
\label{eq:RHP_normalized}
    \overline{\mathcal{N}}_{\mathrm{RHP}} := \frac{1}{N} \sum_{i=0}^{N - 1}\tanh{[h(t_i)^+]}
\end{equation}
The quantity $\overline{\mathcal{N}}_{\mathrm{RHP}}$ belongs to the interval $[0,1)$ and incorporates both the magnitude of the violation of complete positivity and the fraction of the observation interval over which such violations occur. If $h^{+}(t_i)=0$ for every sampled interval, Eq.~\eqref{eq:RHP_normalized}
gives $\overline{\mathcal{N}}_{\mathrm{RHP}}=0$. 

\section{Cross-sections of the reachable population simplex}
\label{ap:cones}

For a four-level system whose state is diagonal in the energy eigenbasis, the state is completely described by the population vector
\begin{equation}
    \mathbf{p}=(p_0,p_1,p_2,p_3)^{\mathsf T},
    \quad p_i\geq 0,\quad\sum_{i=0}^{3}p_i=1.
\end{equation}
Consequently, the set of all physical population vectors is the three-dimensional probability simplex
\begin{equation}
    \Delta_3=\left\{\mathbf{p}\in\mathbb{R}^{4}_{+}\,\middle|\,\sum_{i=0}^{3}p_i=1\right\}.
\end{equation}
This simplex can be represented geometrically as a regular tetrahedron. To this end, we associate the four pure population states with the vertices
\begin{equation}
\begin{split}
    \mathbf{v}_0&=(0,0,0),\\
    \mathbf{v}_1&=(1,0,0),\\
    \mathbf{v}_2&=\left(\frac{1}{2},\frac{\sqrt{3}}{2},0\right),\\
    \mathbf{v}_3&=\left(
        \frac{1}{2},
        \frac{\sqrt{3}}{6},
        \frac{\sqrt{6}}{3}
    \right).
\end{split}
\end{equation}
An arbitrary population vector is then embedded into the tetrahedron through its barycentric coordinates,
\begin{equation}
    \mathbf{r}(\mathbf{p})=
    \sum_{i=0}^{3}p_i\mathbf{v}_i.
\end{equation}
For a given class of dynamics $X$, let
\begin{equation}
    \mathcal{S}_{X}=
    \left\{
        \mathbf{p}(t,\boldsymbol{\theta})\,\middle|\,t\in[0, t_{\max}], \ \boldsymbol{\theta}\in\Theta_X\right\}
\end{equation}
denote the set of sampled population vectors, where $\boldsymbol{\theta}$ represents the dynamical parameters.
The corresponding convex reachable sets are
\begin{equation}
    \mathcal{C}_{X}=
    \operatorname{conv}\left(\mathcal{S}_{X}\right).
\end{equation}
The sampled points in $\mathcal{S}_{X}$ are states directly reached at particular times and for particular choices of the dynamical parameters. By contrast, the interior points introduced by convexification correspond to classical probabilistic mixtures of different dynamical protocols. If such randomization is not regarded as an allowed operation, the convex hull should instead be interpreted as an envelope of the numerically sampled reachable set.

\subsection{Sections at fixed target population.}

To investigate how a prescribed target population can be achieved, we fix
\begin{equation}
    p_2=\gamma,\qquad 0\leq \gamma<1,
\end{equation}
and consider the section
\begin{equation}
    \mathcal{C}_{X}^{\gamma}=\mathcal{C}_{X}\cap\left\{\mathbf{p}\in\Delta_3\,\middle|\,p_2=\gamma\right\}.
\end{equation}
The complete physical simplex at fixed $p_2=\gamma$ is a two-dimensional triangle, whereas $\mathcal{C}_{X}^\gamma$ is generally a convex polygon contained within that triangle. Depending on the dynamics and on the value of $\gamma$, the section may also reduce to a line, a single point, or the empty set.

Once $p_2=\gamma$ is fixed, the remaining populations satisfy
\begin{equation}
    p_0+p_1+p_3=1-\gamma.
\end{equation}
It is therefore convenient to introduce the normalized residual populations
\begin{equation}
   \tilde{p}_j=\frac{p_j}{1-\gamma},\qquad j\in\{0,1,3\}.
\end{equation}
They satisfy
\begin{equation}
    \tilde{p}_0+\tilde{p}_1+\tilde{p}_3=1,
    \qquad
   \tilde{p}_j\geq0,
\end{equation}
and hence define a two-dimensional probability simplex. In the population picture, $\tilde{p}_j$ can be interpreted as the conditional probability of occupying level $j$, conditioned on the system not occupying the target level:
\begin{equation}
   \tilde{p}_j=\Pr\!\left(j\,\middle|\,j\neq2\right).
\end{equation}
The transformation is invertible for each fixed $\gamma$, since
\begin{equation}
    p_2=\gamma,\quad p_j=(1-\gamma)\tilde{p}_j,\quad j\in\{0,1,3\}.
\end{equation}
Thus, no population information is lost by using the variables $\tilde{p}_j$.

The conditional population vector $\tilde{\mathbf{p}}=(\tilde{p}_0,\tilde{p}_1,\tilde{p}_3)$ can be embedded into a regular two-dimensional simplex by choosing
\begin{equation}
    \mathbf{u}_0=(0,0),\quad\mathbf{u}_1=(1,0),
    \quad\mathbf{u}_3=\left(\frac{1}{2},
        \frac{\sqrt{3}}{2}\right),
\end{equation}
and defining
\begin{equation}
    \mathbf{x}(\tilde{\mathbf{p}})
=\tilde{p}_0\mathbf{u}_0+\tilde{p}_1\mathbf{u}_1+\tilde{p}_3\mathbf{u}_3.
\end{equation}
The vertices of this triangle represent configurations in which all the residual population $1-\gamma$ occupies one of the levels $0, 1$ or $3$.

The polygon $\mathcal{C}_{X}^\gamma$ is obtained by intersecting the edges of the three-dimensional convex hull $\mathcal{C}_{X}$ with the plane $p_2=\gamma$. Consider an edge connecting two hull vertices $\mathbf{p}^{(a)}$ and $\mathbf{p}^{(b)}$. Points along this edge have the form
\begin{equation}
    \mathbf{p}(s)=(1-s)\mathbf{p}^{(a)}+ s\mathbf{p}^{(b)},\quad 0\leq s\leq1.
\end{equation}
If the edge crosses the plane $p_2=\gamma$, the intersection parameter is
\begin{equation}
    s=\frac{\gamma-p_2^{(a)}}{ p_2^{(b)}-p_2^{(a)} }.
\end{equation}
Collecting all such intersections gives the vertices of the section polygon. These vertices are subsequently mapped to the two-dimensional $\tilde{\mathbf{p}}$-simplex.

Let the ordered vertices of the resulting polygon be
\begin{equation}
    \mathbf{x}_k=(x_k,y_k),
    \qquad
    k=1,\ldots,N,
\end{equation}
with $\mathbf{x}_{N+1}=\mathbf{x}_1$. Its area can be evaluated using the polygon formula \cite{Braden1986}
\begin{equation}
    A_X(\gamma)=\frac{1}{2}
    \left|\sum_{k=1}^{N}\left( x_k y_{k+1}- x_{k+1}y_k\right) \right|.
\end{equation}
Numerically, the two-dimensional convex hull of the intersection points is constructed, and its polygonal area is evaluated. In the \texttt{scipy.spatial.ConvexHull} convention, this quantity is returned by \texttt{hull.volume}. For a two-dimensional hull, \texttt{hull.area} instead denotes the perimeter.

To quantify how much of the region allowed by thermal operations is reproduced by each dynamical class, we define the TO coverage
\begin{equation}
\eta_X(\gamma,q)=\frac{A_X(\gamma,q)}{A_{\mathrm{TO}}(\gamma,q)},
\quad X\in\{\mathrm{M},\mathrm{NM}\}.
\end{equation}
Thus, $100\,\eta_X(\gamma,q)$ is the percentage of the TO-allowed section covered by the corresponding dynamical hull. The increase in TO coverage obtained by including the non-Markovian trajectories is
\begin{align}
\Delta\eta(\gamma,q) &=
\eta_{\mathrm{NM}}(\gamma,q)-\eta_{\mathrm{M}}(\gamma,q)\\
&=\frac{A_{\mathrm{NM}}(\gamma,q)-A_{\mathrm{M}}(\gamma,q)}{A_{\mathrm{TO}}(\gamma,q)}.
\end{align}

\end{appendix}

\end{document}